\documentclass[utf8]{FrontiersinHarvard}
\usepackage{url,hyperref,lineno,microtype,subcaption}
\usepackage[onehalfspacing]{setspace}

\def\keyFont{\fontsize{8}{11}\helveticabold }
\def\firstAuthorLast{Becher {et~al.}} %use et al only if is more than 1 author
\def\Authors{Ori Becher\,$^{1}$,  Mira Marcus-Kalish\,$^{1}$ and David M. Steinberg\,$^{1}$}
\def\Address{$^{1}$Department of Statistics and Operations Research, Tel Aviv University, Tel Aviv, Israel}
\def\corrAuthor{David M. Steinberg}
\def\corrEmail{dms@tauex.tau.ac.il}

\usepackage[utf8]{inputenc}
\usepackage{graphicx}
\usepackage[a4paper,nomarginpar,bindingoffset=6mm]{geometry}

\usepackage{fancyhdr}
\usepackage{amssymb}
\usepackage{amsmath}
\usepackage{algpseudocode}
\usepackage{booktabs,makecell}
\usepackage{algorithm}
\usepackage{units}
\usepackage{subcaption}
\usepackage{nameref}
\usepackage{booktabs}
\usepackage{xcolor}
\usepackage{comment}
\usepackage{lscape}
\usepackage[compact]{titlesec}
\titlespacing{\section}{0pt}{*0}{*0}
\usepackage{pdfpages}

\begin{document}

\onecolumn
\firstpage{1}

\title {Federated Statistical Analysis: Non-parametric Testing and Quantile Estimation}

\author[\firstAuthorLast ]{\Authors} %This field will be automatically populated
\address{}
\correspondance{}

\extraAuth{}

\maketitle
\bibliographystyle{Frontiers-Harvard.bst}

\begin{abstract}

The age of big data has fueled expectations for accelerating learning.
The availability of large data sets enables researchers to achieve more
powerful statistical analyses and enhances the reliability of conclusions, which can be based on a broad collection of
subjects.
Often such data sets can be assembled only with access to diverse sources; for example, medical research that combines data from multiple centers in a federated analysis.
However these hopes must be balanced against data privacy concerns, which hinder sharing raw data among centers.
Consequently, federated analyses typically resort to sharing data summaries from each center.
The limitation to summaries carries the risk that it will impair the efficiency of statistical analysis procedures.

In this work we take a close look at the effects of federated analysis on two very basic problems, nonparametric comparison of two groups and quantile estimation to describe the corresponding distributions.
We also propose a specific privacy-preserving data release policy for federated analysis with the $K$-anonymity criterion, which has been adopted by the Medical Informatics Platform of the European Human Brain Project. Our results show that, for our tasks, there is only a modest loss of statistical efficiency.

\section{}

\tiny
 \keyFont{ \section{Keywords:} Federated analysis, Mann-Whitney test, Medical informatics, Privacy preservation, Information loss} %All article types: you may provide up to 8 keywords; at least 5 are mandatory.
\end{abstract}

\section{Introduction}\label{section:intro}

The ability to analyze large sets of medical data has clear potential for improving health care. Often, though, a large patient base is available only by combining data from multiple silos. Combining data faces immediate challenges: data quality is often not uniform, nor is granularity, sites may code data differently, requiring adjustment, before analysis is possible. Additionally, given the personal and sensitive nature of medical information, sharing data across centers poses ethical and legal concerns related to privacy preservation. Many countries have enacted laws protecting privacy. For example, data sharing in Europe must be consistent with the General Data Protection Regulation (``GDPR") and in the United States with the Health Insurance Portability and Accountability Act (``HIPAA").

Federated data analysis addresses the privacy concerns by limiting data release from a center to summary statistics, without revealing the raw data. The analysis must then rely on the summary statistics. Methods for federated analysis have been proposed in the machine learning literature, but little research has been done to examine the consequences of the methods for statistical inference.
Our goal in this paper is to fill some of the gap, comparing federated approaches for some basic statistical analyses. A simple example may help to set the stage. The Kaplan-Meier estimator is one of the most widely-used tools in the analysis of medical data. The resulting survival curves show the event times of all subjects and thus compromise privacy.

The roots of our work are in the European Human Brain Project (``HBP"). Data sharing is a major priority for the HBP, but must be fully consistent with the GDPR. Salles et al. (2018) spelled out a detailed Opinion and Action Plan on `Data Protection and Privacy' for the HBP. The plan gives important guidelines and a sound administrative framework for data protection, but does not provide technical solutions. One such solution was adopted in the design of the Medical Informatics Platform (``MIP"), the HBP vehicle for federated, multi-institutional, data analysis. Specifically, any data table exported from a member institution for use in federated analysis on the MIP must have at least 10 subjects in any cell of the table. The Kaplan-Meier curve, cited earlier, requires for each event time a table with the size of the risk set and the number who had an event; the latter will usually be less than 10 and so will not satisfy the privacy constraint.

In section \ref{section:binningAlgorithms} we propose a method for data summary that works specifically with the MIP restriction, generating tables in a federated manner.
We then address two particular statistical problems: (i) use of the nonparametric Mann-Whitney U test (henceforth ``MWU") \cite{MWU} to test the hypothesis that there is no difference between two groups and (in section ~\ref{section:testing}) (ii) quantile estimation to describe the corresponding distributions (in section ~\ref{section:estimation}).
Discussion and conclusions are in section ~\ref{section:summary}.

\section{Related Work}\label{section:relatedWork}

Most of the research on federated data analysis has focused on algorithmic issues, under the general header of Federated Learning.
These works emphasize aspects of
computation and communication efficiency, security, and the adjustment of machine learning algorithms to federated
settings. Several recent surveys provide good summaries
\citep{yang2019}\citep{Li_2021}\citep{Kairouz}. Notable examples are \citep{McMahan}, who presented the ``FederatedAveraging'' algorithm, which combines local stochastic gradient descent on each
client with a server that performs model averaging and ``FedProx'' suggested by \citep{li19}, which deals also with
heterogeneous data.

Nasirigerdeh et al. \citep{gwas} created sPLINK, a system used to conduct Genome-Wide Association studies in a federated manner
while respecting privacy. Algorithms such as linear and logistic regression were adjusted to the federated setting using
data summaries from different data centers.
Duan et al. \citep{Duan},\citep{Duan19} presented privacy-preserving distributed algorithms (``ODAL" and ``ODAL2") to perform
logistic regression. With a focus on efficient communication, they made these \textit{one-shot algorithms}, i.e. using only one information transfer from each center; by contrast, most algorithms are iterative and require multiple transfers. Liu and Ihler \citep{liu2014distributed} considered federated maximum likelihood estimation for parameters in exponential
family distribution models. Their idea was to combine local maximum likelihood estimates by minimizing
the Kullback-Leibler divergence. Their method yields a federated estimator that outperforms any other linear
combination in various scenarios and is equivalent to the global MLE when the underlying distribution
belongs to the full exponential family.

Some related statistical literature is concerned with distributed computing, in which the data is
centralized but so large that calculations are split over multiple servers in parallel to accelerate calculations.
For example, Rosenblatt and Nadler \citep{Nadler} showed that the estimator from averaging estimates from $m$
servers is as accurate as the centralized solution when the number of parameters $p$ is fixed and the amount of data $n\rightarrow \infty$.

Many data sets contain sensitive private information that must legally and ethically remain unexposed.
Several measures of privacy have been proposed, one of which is the degree of anonymization - the extent
to which one is able to identify an individual from the records in the data and link the sensitive information to her.
One well-known criterion for anonymization is $K$-anonymity \citep{kAnonymity}. A dataset is $K$-anonymous if each data
item cannot be distinguished from at least $K-1$ other data items. Fulfilling this criterion introduces fuzziness into
the data that makes it less likely to expose a certain individual. Another popular criterion is differential privacy,
in which querying a data-base must not reveal too much information about a specific
individual's record in it ~\citep{dwork2006}.

One of the techniques to achieve $K$-anonymity is generalization. For example, one could release that $K$ patients were
between age 10 and 30 instead of releasing the ages of each of these patients. This appears to be the motivation for the
privacy rule of the MIP, mentioned earlier, that any data table exported from a member institution for use in federated
analysis must have at least 10 subjects in any cell of the table.

\section{The Binning Algorithm}\label{section:binningAlgorithms}
This section describes a procedure for constructing a $K$-anonymous federated summary table when two groups are compared with respect to a numerical variable. We denote the groups by $x$ and $y$ and use the terms control and treatment for them.
The summary table will have $B$ bins, with the $b$th bin given by $(c_{b-1},c_b]$, and observation frequencies $f_{bx}$ for
the control group and $f_{by}$ for the treatment group. The table preserves $K$-anonymity in that it
is constructed from frequency tables released from the centers in which all cell counts are either 0 or are $\geq K$.

Here is an outline of our table construction process.
We proceed sequentially to add information from each center, beginning with the largest center and proceeding in decreasing order of sample size.
The initial summary table
meets the cell count constraint while attempting to minimize the width of the cells. Data from the other
centers are then added, generating new bins if it is possible to do so without violating the
privacy constraint. Existing bins are never removed. When cell counts from a new center are between $0$ and $K$, neighboring bins
are combined and their total count is redistributed among the bins that were combined
(See Algorithm ~\ref{alg:BinningAlgorithm} for details).

\subsection{Binning the largest center}\label{sections:single center}

The process proceeds (arbitrarily) from small to large values. The first bin is, initially, from
$a_0$ to $a_1$, where $a_0$ is the minimal value in the data and $a_1$, is the smallest data value for which $[a_o,a_1]$
has at least $K$ observations from one group and either 0 or at least $K$ observations from the other group. The next tentative bin
limit, $a_2$, is found in the same way, looking at the interval $(a_1,a_2]$. This continues so long as a new bin limit
can be found. When a limit cannot be found, the number of unbinned data in at least one group is
between $0$ and $K$. Tentatively extend the upper limit of the previously formed bin to the maximal value of this group
as the next limit. The unbinned data from the other group might permit continuation of the process, blocking off new
bins in which that group has counts of at least $K$, versus counts of 0 for the first group. When that group has fewer
than $K$ unbinned data, replace the last bin limit by the maximal value in the second group.
(See Algorithm ~\ref{alg:next2dPoint} in the Appendix for details.)

The initial bin boundaries $a_0,\ldots,a_B$ produced by the algorithm above are actual data values and, unless
many subjects share the same value, violate the privacy condition. There is a simple fix for $a_1,\ldots,a_{B-1}$.
All values in the $j$th bin are $\leq a_j$ and all values in the $j+1$st bin are $> a_j$. So we can replace $a_j$ by
$c_j = w a_j + (1-w)v_{j+1}$ where $v_{j+1}$ is the smallest value in the $(j+1)$st bin and $w$ is a uniform random variable on $(0,1)$. The extreme boundaries $a_0$ and $a_B$ are the minimum and maximum in the data, so a different
approach is needed. One option is to take $c_0 = -\infty$ and $c_B = \infty$. 
Another option is to impose natural limits; for example, if by definition a variable cannot assume negative values, we could choose $c_0 = 0$. A final option is to extend the bin limits by ``privacy buffers''. To make
these reasonably close to the data, we base them on the observed gaps between successive observations in the extreme bin. For
example, compute $c_B$ as $a_B + \bar{d}_{B}$, where $\bar{d}_{B}$ is the mean difference between consecutive data
points in the last bin. (If $\bar{d}_B = 0$, $c_B = a_B$, but this is now privacy preserving, as all observations in the
last bin are equal to one another, with more than $K$ in each group that has data.) Similarly, compute $c_0$ as $a_0 - \bar{d}_{1}$.

\newpage

\textbf{Pseudo code}

\begin{algorithm}
    \caption{Binning algorithm}\label{alg:BinningAlgorithm}
    \begin{algorithmic}
        \State Input: x1, x2
        \State bins, frequencies1, frequencies2 = empty list
        \State
        \While{true}
            \State next\_point = next\_2d\_point(x1, x2)
            \If{next\_point is None}
                \State frequencies1[length(frequencies1)] += length(x1)
                \State frequencies2[length(frequencies2)] += length(x2)
                \State bins[length(bins)] = $\infty$
                \State return bins, frequencies1, frequencies2
            \EndIf
            \State f1 = length(x1[x1$<$next\_point])
            \State x1 = x1[x1 $\geq$ next\_point]
            \State f2 = length(x2[x2$<$next\_point])
            \State x2 = x2[x2 $\geq$ next\_point]
            \State bins.append(anonymize\_boundary(next\_point, x1, x2))
            \State frequencies1.append(f1)
            \State frequencies2.append(f2)
        \EndWhile
        \State{Output\{bins, f1, f2\}}
    \end{algorithmic}
\end{algorithm}

\clearpage

\subsection{Joining additional centers}

A new algorithm is needed to add the data from a new center, preserving all bin boundaries
from the first center. The simple option of
increasing the frequency counts in each current bin is not an option, as the incremental table from the new center will typically not be $K$-anonymous.
Further, the incremental counts for some existing bin might be so large that data from the new center could actually be used to split it into two or more bins.

Algorithm~\ref{alg:federatedAlgorithm} is used to add the information from a new center to an existing summary table. We first iterate over the current bins, creating finer bins if possible. Then we remove any counts that are not $K$-anonymous by combining and redistributing data from adjacent cells. Pseudo-code for Algorithm~\ref{alg:federatedAlgorithm}, and for two algorithms called by it, are given in the Appendix.

Splitting an existing bin into two bins forces us to
reallocate the previous frequencies. We do so proportionally to the relative frequencies from the new center. For example, suppose a bin with a current count of 27 for one group is split into two new bins, which have equal counts at the new center.
Then we split the 27 equally to the two new groups, adding 13.5 to each. Note that this procedure can result in counts that are not integers.

After creating new bins wherever possible, we iterate again and fix
bins where the new center has frequencies between 0 and $K$.  Proceeding from bin 1 to bin $B$, these non-private bins are combined with the next bin to the right until all counts from the new center are either 0 or at least $K$. Then the total counts are distributed among the original bins proportionally to the relative frequencies of the bins in the current table. Table 4 in the Appendix shows an example that illustrates how the algorithm works.

\textbf{Joining the last bin}

The extreme bin limits $c_0$ and $c_B$ must be compared with the minimum and maximum values, respectively, in the new
center. If the new center has a more extreme data value, we need to revise these bin limits. We do so by applying the buffer method that was used to find $C_0$ and $c_B$ in the largest
center, but now adding buffers that depend only on the data in the extreme bin from the new center.

\section{Testing}\label{section:testing}

This section considers the problem of hypothesis testing with federated data, studying the common problem of determining whether numerical outcomes from two groups come from the same distribution (the null hypothesis, $H_0$); or whether one group has larger values than the other.
The standard choice is the independent samples $t$-test,
which requires the mean, the standard deviation and the number of observations in each group. All of these are privacy-preserving summary statistics, so the $t$-test can still be used with federated data. However, the $t$-test relies on the assumption, often invalid, that the data are normally distributed. We consider here the standard non-parametric alternative, the Mann-Whitney U (``MWU")
test~\citep{MWU} (or, equivalently, the Wilcoxon rank sum test).

\subsection{The Mann-Whitney U test}\label{subsec:the-mann-whitney-u-test}

The MWU statistic can be defined as follows. Denoting the observations in the two groups by $X_1,\ldots,X_n$ and $Y_1,\ldots,Y_m$,

\[U=\sum_{i=1}^{n} \sum_{j=1}^{m} S\left(X_{i},Y_{j}\right)\]

with \[S(X_i,Y_j)=\left\{\begin{array}{ll}
                             1 & Y_j>X_i \\ 0 & Y_j=X_i \\ -1 & Y_j<X_i
\end{array}\right.\]

If $H_0$ is true, the expected value of $U$ is 0 and its variance is
$V=\frac{m n(N+1)}{3}\left[1-\frac{\sum_{r=1}^{D}\left(t_{r}^{3}-t_{r}\right)}{N\left(N^{2}-1\right)}\right]$, where $N=n+m$, $D$ is the number of distinct values in the data, and $t_r$ is the number of observations that share the $r$th distinct value. The second term corrects the variance for the presence of ties in the data.
If $Y\stackrel{d}{=}c+X ,$ $c\in \mathbb{R}$, the distribution of $U$ is stochastically increasing as a
function of $c$.
The power of the test depends on $P(Y>X)$ and is high when this probability differs from 0.5.

The MWU test involves direct comparison of each data point in one group with each data point from the other group. As this includes comparisons of observations from different centers, it is impossible to
compute the MWU statistic for a federated analysis.
Two broad options are possible for federated  analysis.
\begin{itemize}
    \item Compute the MWU statistic separately for each center and then combine them across centers.
    \item Generate a federated table summarizing the data from all the centers and then compute the MWU statistic on the federated table.
\end{itemize}

The next subsections present options for combining center-specific MWU statistics and the second analysis option, used in conjunction with our federated binning algorithm.

\subsection{Sum of $U$-statistics}\label{subsec:sum-of-$u$-statistics}

Denote by $U_l$ the MWU from the $l$th center, based on $n_l$ and $m_l$ observations from the two groups,
with $N_l=n_l+m_l$; and denote by $V_l$ its variance under $H_0$.
A simple way to form a federated test statistic is to sum the individual statistics over the centers and normalize them
by their standard deviation, leading to
\begin{equation}
    T_{s u m}=\frac{\sum_{l=1}^{L} U_{l}}{\left( \sum_{l=1}^{L} V_{l} \right)^{0.5}}  \stackrel{H_ 0}{\rightarrow } N(0,1).
    \label{eq:sumstatistic}
\end{equation}

\subsection{Weighted average of $U$-statistics}\label{subsec:weighted-average-of-$u$-statistics}

A simple generalization is to replace the sum of the statistics by a weighted sum, with an optimal choice of weights.
It is convenient to do this using the normalized test statistics for each center, $Z_l=U_l/V_l^{0.5}$.
The weighted test statistic is then
\begin{equation}
    T_{w}=\frac{\sum_{l} a_{l} Z_{l}}{\left(\sum_{l} a_{l}^{2}\right)^{0.5}} \stackrel{H_0}{\rightarrow} N\left(0 , 1\right).
    \label{eq:weightedsum}
\end{equation}
The choice of weights can be made to maximize the power of the test when the null hypothesis is not true, using the
fact that
\[T_{w}=\frac{\sum_{l} a_{l} Z_{l}}{\left(\sum_{l} a_{l}^{2}\right)^{.5}} \stackrel{H_1}{\rightarrow} N\left( \frac{\sum_{l} a_{l} \delta_{l}}{\left(\sum_{l} a_{l}^{2}\right)^{.5}}, 1 \right)\]
where $\delta_l$ is the standardized effect in center $l$.
For the MWU statistic, the standardized effect can be expressed as \[\delta_l = E(Z_l)=\frac{m_ln_l(P_l^+-P_l^-)}{SD(U_l|H_0)},\]
where $P_l^+ = P(Y>X)$ and $P_l^- = P(Y<X)$ in center $l$.
Although the formula permits the probability difference to vary over centers, the natural basis for defining the
weighted sum statistic is to assume a constant difference, in which case the optimal weights depend on the sample sizes and, if present, the extent of tied data.
See equation ~\ref{equation:WMWU} in the Appendix for derivation of the weights.

\subsection{Fisher's method}\label{subsec:fisher's-method}

Fisher's method~\citep{Fisher} combines the p-values from independent samples. The corresponding statistic is
$T_{F} = -2\sum_{l=1}^L\log(p_l)\stackrel{H_0}{\rightarrow}  \chi^2_{2L}$ where $p_l$ is the $p$-value from the MWU test result
in the $l$th center.

\subsection{Federated table MWU statistic}\label{subsec:summary-table-mwu-statistic}

We can compute the MWU statistic from the federated summary table generated by the algorithm described in section~\ref{section:binningAlgorithms}.
The table will have $B$ bins whose
frequencies are $fx_i$ and $fy_i$.
The frequencies sum to the total amount of data over all the centers, but need not be integers.

The MWU statistic for the federated table compares observations on the basis of their bins and is given by
\begin{equation}
    U_{fed}=\sum_{i=1}^{B} \sum_{j=1}^{B} f x_{i} f y_{j} S\left(c_{i}, c_{j}\right)
    \label{eq:fedstatistic}
\end{equation}
where  $c_0<c_1<\dots<c_B$ are the endpoints of the bins and
\[S(c_i,c_j)=\left\{\begin{array}{ll}
                        1 & c_j>c_i \\ 0 & c_j=c_i \\ -1 & c_j<c_i
\end{array}\right.\]
The variance of $U_{fed}$ can be computed from the formula in section \ref{subsec:the-mann-whitney-u-test}, keeping in mind that all observations in the same bin are tied.

\subsection{Comparison of the tests}\label{subsec:simulation-results2}

A simulation study was used to compare the different
federated MWU tests to an analysis of the combined data.
Our goals are to assess how the federated analysis affects the power of the tests, and to use the power
analysis to compare the testing methods.
We also vary the simulation settings to examine how the results and comparisons are affected by the
number of centers in the study and by heterogeneity across centers.

We simulated situations with 1500 observations in each group, divided over 3, 5 or 10 centers, with the number of observations unbalanced among the centers
(see Table~\ref{table:nobservation}).

\begin{table}[h]
    \caption{Observations per center}\label{table:nobservation}
    \centering
    \begin{tabular}{|c | c | c|}
        \toprule
        Number of centers & Number of observations in each group         & each group total \\
        \midrule
        3                 & 698, 476, 326                                & 1500             \\
        5                 & 492, 368, 276, 208, 156                      & 1500             \\
        10                & 307, 250, 208, 172, 143, 118, 98, 81, 67, 56 & 1500             \\
        \bottomrule
    \end{tabular}
\end{table}

For the Mann-Whitney test, only the order of the observations is important, so any distribution can be
used to simulate the data. Our model generates control group observations at center $l$ as
\[x_{il} = \epsilon_{il}+ \alpha_{l} \]
and treatment group observations as
\[y_{jl} = \epsilon_{jl} + \alpha_{l}+\beta_{l}.\]

The possibility that centers may differ from one another is represented by
$\alpha_l\sim N(0,\sigma_\alpha^2)$.
The difference between treatment and control at center $l$ is
$\beta_l \sim N(\delta,\sigma_{\beta}^2)$ where $\delta$ is the overall difference, and $\sigma_{\beta}$ represents
heterogeneity of the treatment effect across centers.
The terms $\epsilon_{il},\epsilon_{jl}\sim N(0,1)$ are random errors. All random variables are independent of one another.

We simulated experiments with
several different combinations of input parameters.
We chose $\sigma_\alpha \in \{0,0.1,0.2\}$ and $\sigma_\beta \in \{0,0.05,0.06\}$ to achieve between center variance, and $\delta \in \{0,0.05,0.1\}$.
Including $\delta = 0$ allowed us to verify that the tests remain reliable when both groups have the same mean. Note, however, that the variance is slightly larger for the treatment group if $\sigma_{\beta} > 0$, so that this setting does not fully match the null hypothesis of identical distributions.

Figure S1 (in the supplementary file) shows the distributions of $p$-values for all the tests in the null setting $\delta = 0$. The left panel panel includes heterogeneity across centers ($\sigma_{\alpha}=0.1$), but no effect heterogeneity, and shows a uniform distribution for all the tests, as desired. The right panel adds a small amount of effect heterogeneity ($\sigma_{\beta}=0.05$. This results in a slightly wider spread of $p$-values for all the tests, so that actual type 1 errors are inflated from their nominal values. The fraction of $p$-values below 0.05 (0.01) was approximately 0.08 (0.025). The inflation was slightly weaker when more centers were included and slightly larger only for Fisher's test. The additional bias of Fisher's test is not surprising, as it is sensitive to the existence of an effect within a center, but not to having a consistent direction of the effect.

\begin{figure}[H]
    \includegraphics[width=15cm, height=18cm]{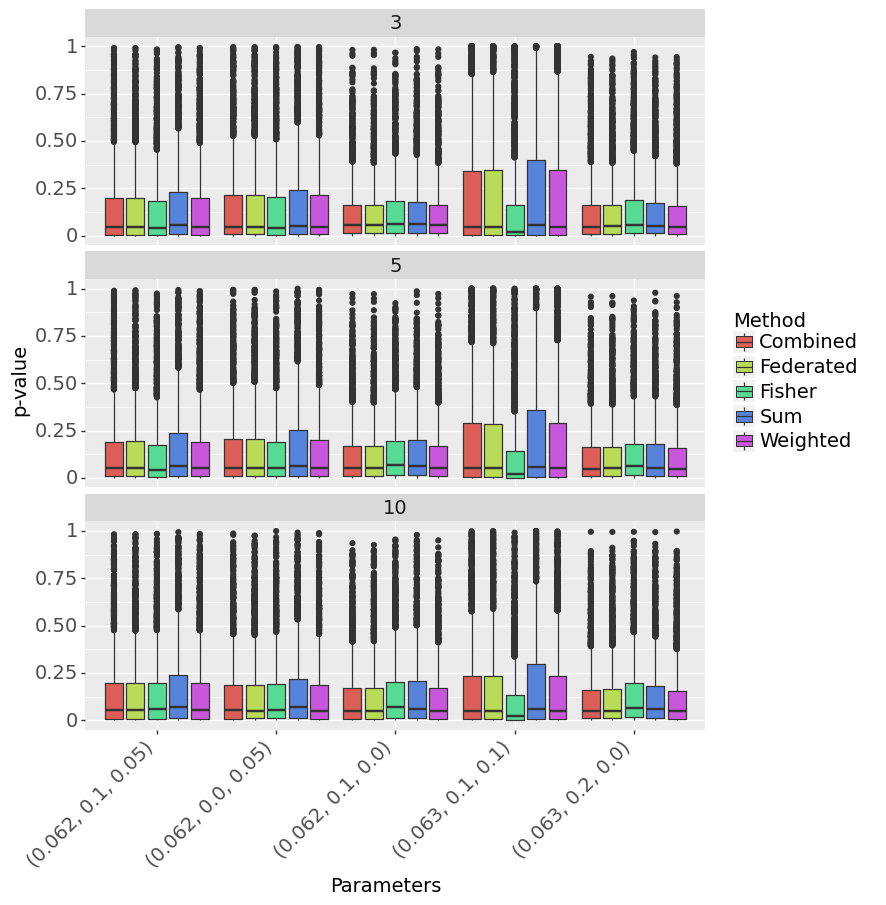}
    \centering
    \caption{Comparison of methods across different parameters and number of centers. Panels represent the number of centers, the $Y$-axis presents the $p$-values and the $X$-axis the parameters $( \delta,\sigma_\alpha,\sigma_\beta)$.
    The different methods are color-coded.}
    \label{fig:pval_by_effect_center}
\end{figure}

Figure~\ref{fig:pval_by_effect_center} compares methods when $\delta \neq 0$ across different
parameters and numbers of centers. See also Supplementary Table S1.
The federated table and weighted tests have $p$-value distributions that are very similar to those from combining all the data, indicating almost no loss of power. The sum test has higher $p$-values, hence consistently lower power. The $p$-values with Fisher's method are a bit higher when the treatment effect is consistent across centers ($\sigma_{\beta}=0)$. When the effect is not consistent, they are lower. However, as already seen, Fisher's test in this case fails to preserve type 1 error, with a bias toward low values.

\begin{figure}[H]

    \includegraphics[scale=0.25]{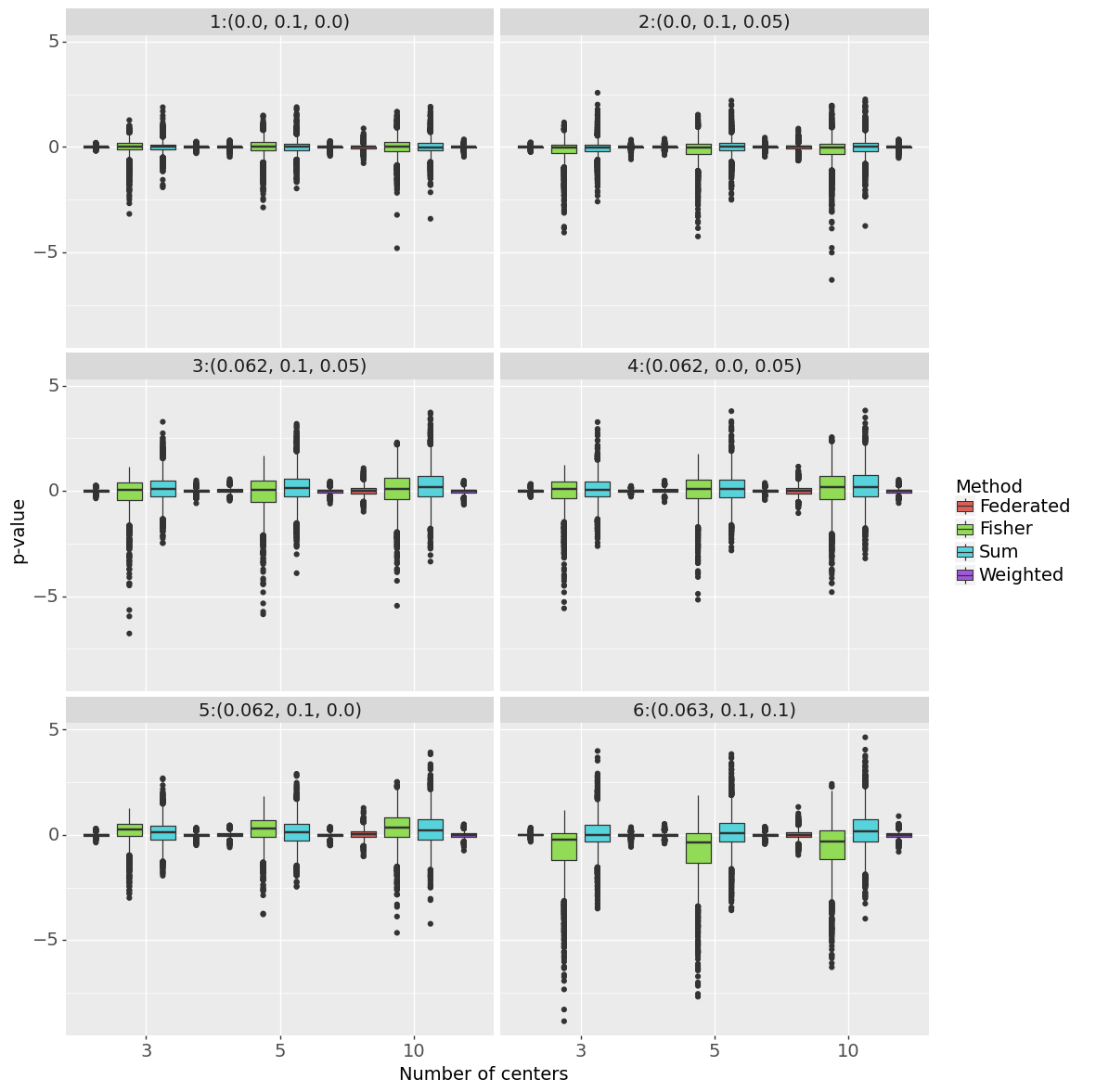}
    \centering
    \caption{The figure shows $log(p_{iv}/p_{is})$ on the $Y$ axis, where $s$ is the combined data analysis.
    The panels correspond to the different parameter settings for $(\delta,\sigma_\alpha,\sigma_\beta)$.
    The number of centers is on the $X$ axis and the methods are color-coded.}
    \label{fig:pvals_boxplots}
\end{figure}

Figure~\ref{fig:pvals_boxplots} focuses on how closely the federated test results compare with those from the combined
test (i.e. using the full data) by comparing the $p$-values of each method on the same simulated data set. The $Y$ axis presents  $log(p_{iv}/p_{is})$ where  $i$ represents the simulation number, $s$ is the combined test and $v$ is the federated test.

Across all the settings, the weighted test most closely replicates the $p$-value of the combined test. The federated table is also similar, but more variable, especially when $\delta \neq 0$.
In the top left panel, where $H_0$ is true, all methods are
similar to the combined test. However, adding treatment heterogeneity
(top right panel) induces negative bias in the $p$-values from Fisher's test and increases the variance of the log ratio for that test and for the sum.
In all the settings with center heterogeneity ($\sigma_{\alpha} > 0$), the sum test gave, typically, slightly higher $p$-values than the combined test, hence had lower power.

\begin{figure}[h]
    \includegraphics[scale=.3]{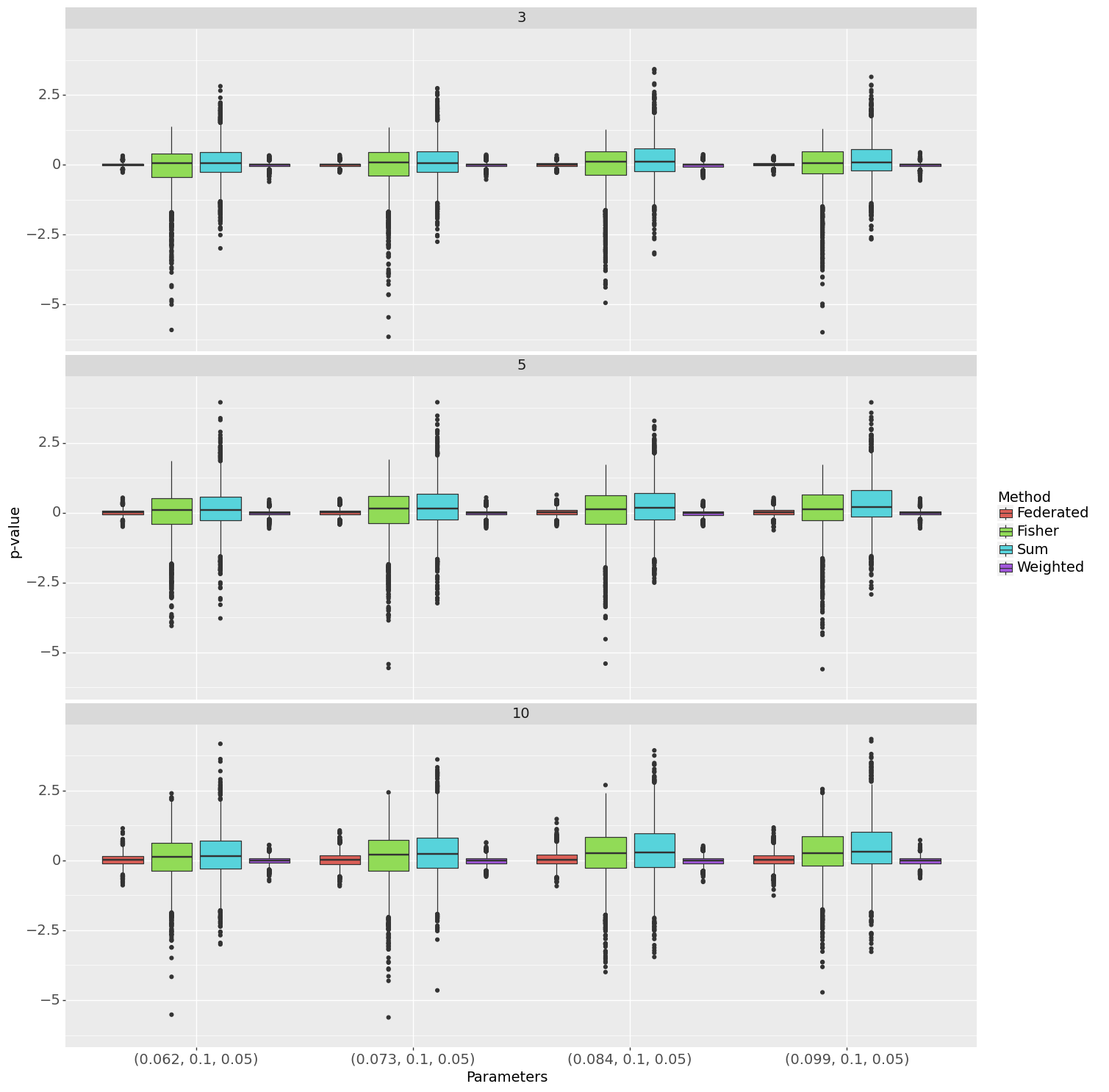}
    \centering
    \caption{Comparison of methods across different parameters and number of centers. Rows correspond to the number of centers and segments within rows
    represent hyper-parameter configurations. The $Y$-axis represents the $p$-values.
    The methods are color-coded.}
    \label{fig:power_by_effect_center}
\end{figure}

To assess the power of the tests as a function of the effect size, we measured the $p$-values over a set of 4
increasing values of $\delta$, when $\sigma_{\alpha}=0.1$ and $\sigma_{\beta}=0.05$.
Figure ~\ref{fig:power_by_effect_center} compares
the methods to the unconstrained test using $\log(p_{iv}/p_{is})$ ($Y$-axis) where  $i$ represents the simulation number, $s$ is the unconstrained method and $v$ is the other method. Again the weighted test is most similar to the combined test, followed by the federated table. Table ~\ref{tab:pavlues_power} shows quantiles of the $p$-value distributions with 10 centers. The quantiles for the weighted test are consistently the lowest ones; with even the modest heterogeneity present here, they are lower even than those for the combined test. Similar quantiles were found for 3 and for 5 centers, indicating that, for the settings we examined, the number of centers has little effect on power.

\clearpage

\section{Estimation}\label{section:estimation}

This section considers the problem of quantile estimation when data are located in different centers.
Quantile estimates are valuable for directing visual summaries of data distributions such as histograms or
Kaplan-Meier plots.
Standard methods for computing sample quantiles cannot be used, as they begin by ordering all the data, violating privacy.
We propose and compare several methods for federated quantile estimation. Throughout we denote by $F(x)$ the CDF and by $Q_p = F^{-1}(p)$ the $p$th quantile of the distribution.

\subsection{Federated estimates using the quantile loss}\label{
    sec:estimating-quantiles-from-federated-data-using-the-quantile-loss
}

A quantile can be estimated as the solution to a minimization problem,
\begin{equation}
        \hat{Q}_{p,Loss}=\underset{q }{\arg \min }\left[(p-1) \sum_{y_{i}<q}\left(y_{i}-q\right)+p \sum_{y_{i} \geq q}\left(y_{i}-q\right)\right],
\label{eq:quantile_loss}
\end{equation}
where the target function is the \textit{quantile loss function}.
The optimization can be carried out on federated data by returning function and gradient values from each center, proceeding iteratively to compute $\hat{Q}_{p,Loss}$. The need for an iterative algorithm to minimize the loss, has the drawback of communication inefficiency.

A more serious concern is that the quantile loss compromises privacy. The loss function within each center is piecewise linear with a change in derivative at each data value in the center. Thus the information from a collection of calls can be used to recover the original data values at the federated node.

Despite the privacy violation, we will include $\hat{Q}_{p,Loss}$ in the subsequent comparisons as a benchmark.

It is possible to exploit the loss function to compute approximate quantile estimators that are differentially private (~\citep{Kaplan22}).

\subsection{Estimating quantiles from the federated data using the Yeo-Johnson transformation}\label{sec:estimating-quantiles-from-federated-data-using-the-yeo-johnson-transformation-directly-over-the-raw-data}

The binning algorithm we introduced in Section \ref{section:binningAlgorithms} can be used to compute a federated estimate of $Q_p$ that is $K$-anonymous.
Here we apply the single group version of the algorithm which gives a summary table that has $B$ bins with endpoints $b_0<b_1<b_2<\dots<b_B$ and frequencies
$f_{x,k}$. Let $\hat{F}_{x,i}$ denote the cumulative
distribution for the federated table at $b_i$.

A naive estimate is the smallest bin limit with cumulative frequency greater than $100p\%$
of the data. However, restricting $Q_p$ to the set of bin limits is an obvious drawback, especially for quantiles in the tails of the distribution. A simple improvement is to interpolate the estimated CDF from one bin limit to the next. Linear interpolation corresponds to the assumption of a uniform distribution within each bin. That may be reasonable for bins in the center of the data. Howver, it is not likely to work well in the tails, especially in the most extreme bins. We did attempt to use linear interpolation, but the results were poor and are not reported here.

We propose here a more sophisticated interpolation method based
on the Yeo-Johnson transformation (``YJ'')~\citep{yj},
a power transformation used to achieve a distribution that is closer to the normal.
The approach extends the well-known Box Cox~\citep{boxcox} transformation to also handle variables that can take on
negative values. The transformation is defined by
\begin{equation}
    h_{\lambda}(x)=\left\{\begin{array}{ll}
                              \left((1+x)^{\lambda}-1\right) / \lambda       & \text { if } \lambda \neq 0 \text { and } x \geq 0 \\
                              \log (1+x)                                     & \text { if } \lambda=0 \text { and } x \geq 0      \\
                              -\left((1-x)^{2-\lambda}-1\right) /(2-\lambda) & \text { if } \lambda \neq 2 \text { and } x<0      \\
                              -\log (1-x)                                    & \text { if } \lambda=2 \text { and } x<0
    \end{array}\right.
    \label{equasion:YJ}
\end{equation}

\paragraph{YJ Table method}\label{methods:yj_table}
In this method the goal is to find values of $\lambda,a_0$ and $a_1$ for which the transformed bin limits
approximately match a normal distribution with mean $a_0$ and standard deviation $a_1$,
\begin{equation}
    h_{\lambda}(b_k) \approx a_0+a_1\Phi^{-1}(\hat{F}(b_k))
    \label{eq:density_model}
\end{equation}
where $b_k$ is a bin limit,
$\hat{F}$ is the estimator of the distribution function from the federated table and $h_{\lambda}(x)$ is the (``YJ")~\citep{yj} transformation.
The quantile $Q_p$ is then estimated by
\begin{equation}
\hat{Q}_{p,YJTable}= h^{-1}_{\hat{\lambda}}(\hat{a}_0+\hat{a}_1\Phi^{-1}(p)).
\end{equation}

Given $\lambda$, we can compute $a_0, a_1$ using linear regression.
To estimate $\lambda$, we use the idea that an effective transformation $h_{\lambda}$ should have transformed quantiles that are linearly related to the YJ-estimated quantiles.
This can be achieved by choosing $\lambda$ to maximize the
correlation between them,
\begin{equation}
    \hat{\lambda}=\underset{\lambda}{\operatorname{argmax}}\left(\operatorname{cor}\left(\Phi^{-1}(\hat{F}(X)), h_{\lambda}( X)\right)\right)
    \label{eq:table_lambda}
\end{equation}
where the values of $X$ we use are the interior bin limits $b_1,\ldots,b_{K-1}$.

Note that the range of the inverse transformation in~\ref{equasion:YJ} is given by
\begin{equation}
    h_{\lambda}(\mathbb{R})=\left\{\begin{array}{ll}
    (-1 /|\lambda-2|, \infty)
                                       & \text { if } \lambda>2             \\
                                       \mathbb{R}              & \text { if } 0 \leq \lambda \leq 2 \\
                                       (-\infty, 1 /|\lambda|) & \text { if } \lambda<0
    \end{array}\right.
    \label{eq:YJ inverse}
\end{equation}

To ensure that the inverse transformation has values in $\mathbb{R}$ we set the constraint
$0\leq \lambda \leq 2$ for equation ~\ref{eq:table_lambda}.

The YJ Table method is a ``one pass'' algorithm, calling the data only to produce the federated summary table. Thus it enjoys full communication efficiency.

\paragraph{YJ Likelihood method}\label{methods:yj_table}The parameters in the YJ transformation can also be estimated by maximum likelihood. Denoting by $x_{il}$ the observations from center $l$ and
by N the total number of observations, the log likelihood is
\begin{equation}
    -N / 2 \log \left(\hat{\sigma}^{2}_\lambda\right)+(\lambda-1) \sum_{l=1}^L\sum_{i=1}^{n_l}
    \operatorname{sign}\left(x_{il}\right) \log \left(\left|x_{il}\right|+1\right)\label{eq:yj_log_likelihood}
\end{equation}
where
\begin{equation}
    \hat{\sigma}^{2}_\lambda = \frac{1}{N}\sum_{l=1}^L\sum_{i=1}^{n_l}h_{\lambda}(x_{il})^2 -
    (\frac{1}{N}\sum_{l=1}^L\sum_{i=1}^{n_l}h_{\lambda}(x_{il}))^2.
    \label{eq:yj_data_log_likelihood}
\end{equation}
For a fixed value of $\lambda$, the log-likelihood requires only
summary statistics
from each center, so can be computed in a federated manner.
This can be embedded in a simple optimization routine that maximizes the log likelihood over $\lambda$.

As with the quantile loss, the YJ likelihood method employs an iterative algorithm, and thus is not communication efficient. However, unlike the quantile loss, the YJ log likelihood for each center is not a simple function of the data that can be immediately inverted to recover data values. Thus the privacy violations of the quantile loss do not occur here.

Once we have $\hat{\lambda}$, we can again use summary statistics from the centers to compute  $\hat{\mu}_{\hat{\lambda}}, \hat{\sigma}_{\hat{\lambda}}$.
The resulting quantile estimator is
\begin{equation}
    \label{eq:YJData}
    \hat{Q}_{p,YJData} = h^{-1}_{\hat{\lambda}}(\hat{\mu}_{\hat{\lambda}}+\hat{\sigma}_{\hat{\lambda}}\Phi^{-1}(p)).
\end{equation}

The likelihood maximization is iterative, so requires multiple communication steps with each center. By contrast, the methods based on the federated table are ``one pass'', requiring just one call to each center. This communication inefficiency of the maximum likelihood method can be improved by
submitting to each center a grid of possible $\lambda$ values.
The centers then return the moments needed to compute the log likelihood for each value in the grid.
The resulting estimate of $\lambda$ can either be the best value among those in the grid or the maximizer of an empirical fit to the relationship between the log likelihood and $\lambda$. The result is an approximate, one pass MLE.

\subsection{Constructing summary tables from quantile estimates}\label{sec:constructing-summary-tables-out-of-quantile-estimates}

Federated quantile estimates can be used to generate an alternative summary table, which presents a collection of quantiles. See table~\ref{tab:quantiles_based_table} for an example, with estimates from optimizing the quantile loss and the $YJ$ likelihood.

\subsection{Simulation Results}\label{sec:simulation-results}

We compared the three quantile estimators using
a simulation configuration similar to that in the testing chapter. As the quantiles are univariate summaries, we
generated data and estimated quantiles only in one group. Another important difference is that the form of the underlying distribution affects the estimation results. In particular, methods may vary when faced with long rather than short tails. To gain insight into this issue, we chose the Gamma as the base distribution for assessing
the quality of quantile estimation.

Each simulated data set included 1500 observations, spread across 3, 5 or 10 centers exactly as described in Table \ref{table:nobservation}.
The observations were generated from the following model:  $x_{il} = \epsilon_{il}\exp(\alpha_{l})$ where $x_{il}$ is
observation $i$ at center $l$ with $\alpha_{l}\sim N(0,\sigma_\alpha^2)$
and $ \epsilon_{il}\sim Gamma(r,1)$ $r \in \{4, 10 \}$.
The skewness of Gamma is $\frac{2}{\sqrt{r}}$, so the smaller value for $r$ has a longer right tail.

For the Gamma data, heterogeneity across centers was induced using scale rather than location shifts.
The value of $\sigma_{\alpha}$ was chosen to achieve between center heterogeneity similar in extent to that in section \ref{section:testing}. There the key term was the ratio $\sigma_{\alpha}/\sigma_{\epsilon}$, which was taken to be 0, 0.1 or 0.2. With Gamma data, the standard deviation of the homogeneous data is proportional to the median, so the analogous choice is to set $\sigma_{\alpha} = \log\left(\frac{Q_{0.5}+\phi\sigma_{\epsilon}}{Q_{0.5}}\right)$, with $\phi$ similar to the values chosen above. We used only $\phi = 0.1$ in our simulations for quantile estimation.

For each combination of the parameters, 2000 simulations were run.
The true quantile $Q_p$ for each simulation was computed from
the mixture (over centers) distribution
by solving the equation below with $l$ as
the center index.

\begin{equation}
    \sum_{l=1}^L \frac{n_l}{N}\Gamma_r(x/exp(\alpha_{l}))=p\label{eq:standard_gamma}
\end{equation}
where $N$ is the number of observations from all centers, $n_l$ the observations in center $l$ and $\Gamma_r$ is the
standard Gamma CDF with shape parameter $r$.
A dominant part of the quantile estimation errors is the natural variability of the underlying Gamma distribution. As the
standard deviation for $Gamma(r,1)$ is $\sqrt{r}$, we summarized results via the normalized estimation error   $\left(\hat{Q}_p-Q_p\right)/\sqrt{r}$ where $\hat{Q}_p$ is the estimator of $Q_p$.

The simulation results for estimating $Q_{0.98}$ are shown in Figure~\ref{fig:errs098}. This quantile is presented separately, as it is the most challenging case, in the right tail of a right-skewed distribution. Results for $Q_{0.02}$,$Q_{0.25}$,$Q_{0.5},Q_{0.75}$ are depicted in Figure~\ref{fig:errs_all}. Further detail is provided in Tables S.2, S.3, S.4 and S.5 (supplementary file), which give, respectively, the estimated bias and standard deviation, the mean squared error (MSE), and the ratio of squared bias to variance for all the methods and all the quantiles.

The YJ data estimator achieved lower MSE than the quantile loss estimator. For the extreme quantiles, the decrease in MSE ranged from 14\% to 44\%. The ``one pass'' YJ table estimator was very accurate for estimating the median and the quartiles, but lost efficiency for the extreme quantiles with the more skewed of the two Gamma distributions and when the number of centers was large. In that setting, the estimator for $Q_{0.02}$ suffered from negative bias and its MSE was almost 3 times as large as for the quantile loss estimator; the MSE for $Q_{0.98}$ was about 80\% larger.

For the settings we studied, variance was the dominant component of MSE. Bias was a substantial problem only in a small number of cases. The YJ methods had large positive bias for $Q_{0.98}$ when $r=4$; however, when $r=10$, and the distribution is itself closer to normal, the bias was negligible.

\section{Summary}\label{section:summary}

In this work we presented novel methods for federated data analysis and investigated their statistical properties.
We proposed a simple algorithm for creating $K$-anonymous data tables in one- and two-group problems and we compared federated approaches for the nonparametric Mann-Whitney U (MWU)test and for estimating quantiles.
Our federated data table is created in a ``one pass'' format, so that it is communication efficient.

For the MWU test, we found that the most powerful method was the weighted average of the MWU statistics from the individual centers, with weights reflecting the sample sizes. This statistic is also communication efficient, gives very similar p-values to those from the combined data and has the advantage of adjusting for inter-center heterogeneity, effectively treating each center as a block. The test based on our federated table was less effective. However, the p-value distributions from the table were only slightly worse than those from the combined data and the weighted average, indicating only a small loss of statistical power.

The fully optimized YJ method consistently had the lowest MSE of the methods we compared. For the extreme quantiles, it improved by 14\% to 44\% over the quantile loss estimator. The ``one pass'' YJ table estimator had almost identical MSE for estimating the median and the quartiles, but lost efficiency for the extreme quantiles when the number of centers was large. The increase in MSE was more substantial (almost 80\%) with the more skewed of the two Gamma distributions we studied.
This is not surprising: our method exploits a transformation to normality and is less successful when the distribution is further from the normal.

It is important that research on federated data analysis will relate to statistical efficiency and not just to algorithmic efficiency. Our work opens this avenue, but much more could be done. Here are some examples.
Our construction method for a federated summary table could be extended to
multiple variables and to
higher dimensions; our method creates the bins in a way fitted to a one-dimensional variable.
This would be needed, for example, to produce a federated analogue of a scatter plot.
Our findings suggest that heterogeneity can harm the federated analysis. Methods are needed to identify heterogeneity and to account for it in the analysis. The investigation of quantile estimators could be extended to a wider class of distributions. Our implementation of the YJ method applies a single transformation to the distribution. For quantiles in the tails of the distribution, it might be better to use separate transformations in the left and in the right tails.

\bibliography{bibliography}

\section*{Appendix}

\subsection*{Additional algorithms}

We present here pseudo-code for some of the algorithms used to construct a federated summary table.

The following algorithms identify the next largest value that can serve to mark a bin boundary for, respectively, a two-group and a one-group federated data table.

\begin{algorithm}
\caption{next 2d point}\label{alg:next2dPoint}
\begin{algorithmic}[1]
    \State input\{x1,x2\}
    \State q1 = next\_1d\_point(x1)
    \State q2 = next\_1d\_point(x2)
    \If{not (valid\_point(x2, q2) and valid\_point(x1, q1))}
        \State return None
     \EndIf
    \If{valid\_point(x2, q1) and valid\_point(x1, q2)}
        \State return min(q1, q2)
    \State mx = max(q1, q2)
    \EndIf
    \State return mx

\end{algorithmic}
\end{algorithm}

\begin{algorithm}
\caption{next 1d point}\label{alg:next1dPoint}
\begin{algorithmic}[1]
    \State input\{x\}
    \Comment*{k is the number of the privacy condition}
    \If if length(x) in \{1,...,k\}:
        \State return None
    \EndIf
    \If length(x[k:]) in \{1,...,k\}:
        \State return x[-1]
    \EndIf
    \State return x[k]
\end{algorithmic}
\end{algorithm}

Algorithm \ref{alg:federatedAlgorithm} describes how to add data from a new center to an existing table. It refers to Algorithm \ref{alg:fixFrequencies}, which combines and reallocates data at the new center so that all information released is $K$-anonymous.
In turn, that algorithm calls Algorithm \ref{alg:nextPrivateSsubset}, which identifies the bins that need to be combined to achieve a combined bin that is $K$-anonymous.

\begin{algorithm}
    \caption{Federated Algorithm\label{alg:federatedAlgorithm}}
    \begin{algorithmic}[1]
        \State input\{x1, x2, bins, f1, f2\}
        \State new\_bins, f1\_student, f1\_teacher, f2\_student, f2\_teacher = empty\_lists
        \For{bin\_i in {1,...,length(bins\_s)}}
            \State current\_bin = bins[bin\_i]
            \State f1\_i = f1[bin\_i]
            \State f2\_i = f2[bin\_i]
            \State current\_x1, current\_x2 = x1[x1 $<$ current\_bin], x2[x2 $<$ current\_bin]
            \State x1, x2 = x1[x1 $\geq$ current\_bin], x2[x2 $\geq$ current\_bin]
            \State new\_sub\_bins, new\_f1, new\_f2 = BinningAlgorithm(current\_x1, current\_x2)
            \If{ length(new\_sub\_bins) $<$ 2}
            %# a.extend(c) -> insert c as the last member of a
                \State new\_bins.append(current\_bin)
                \State f1\_student.append(length(current\_x1))
                \State f2\_student.append(length(current\_x2))
                \State f1\_teacher.append(f1\_i)
                \State f2\_teacher.append(f2\_i)
            \Else
                \State new\_sub\_bins[length(new\_sub\_bins)] = current\_bin
                \State new\_bins.extend(new\_sub\_bins)
                %# a.extend(b) -> concatenate a and b
                \State f1\_student.extend(new\_f1)
                \State f2\_student.extend(new\_f2)
                \State f1\_teacher.extend(reallocate(f1\_i,new\_f1))
                \State f2\_teacher.extend(reallocate(f2\_i,new\_f2))
            \EndIf
        \EndFor
        \State f1\_student = fix\_student\_frequencies(f1\_teacher, f1\_student)
        \State f2\_student = fix\_student\_frequencies(f2\_teacher, f2\_student)
        \State Output\{new\_bins, f1\_student + f1\_teacher,f2\_student + f2\_teacher\}
        \Comment{+ here means element-wise sum}

    \end{algorithmic}
\end{algorithm}

\begin{algorithm}
    \caption{fix student frequencies}\label{alg:fixFrequencies}
    \begin{algorithmic}[1]
        \State input\{f\_teacher, f\_student\}
        \State f\_student\_new = empty list
        \While{length(f\_student) $>$ 0}
            \State f\_student\_subset = next\_private\_subset(f\_student)
            \Comment{See (1) below}
            \State subset\_length = length(f\_student\_subset)
            \State f\_teacher\_subset = f\_teacher[:subset\_len]
            \State f\_student\_new.extend(reallocate(sum(f\_student\_subset),f\_teacher\_subset))
            \State f\_student = f\_student[subset\_len:]
            \State f\_teacher = f\_teacher[subset\_len:]
        \EndWhile
        \State  Output\{f\_student\_new\}

    \end{algorithmic}
    (1) next\_private\_subset finds the set of frequencies whose sum is private. see \label{alg:nextPrivateSsubset} in the
    appendix.
\end{algorithm}

\begin{algorithm}
\caption{next private subset}\label{alg:nextPrivateSsubset}
\begin{algorithmic}
    \State Input\{f\_student)\}
    \State f\_student = f\_student[:]
    \State f\_student\_subset = []
    \State f\_student\_subset.append(f\_student.pop(0))
    \While {not valid(sum(f\_student\_subset))) and length(f\_student) $>$ 0}
        \State f\_student\_subset.append(f\_student.pop(0))
    \If{not valid(sum(f\_student))}
        \State f\_student\_subset.extend(f\_student)
    \EndIf
    \EndWhile
    \State Output\{f\_student\_subset\}

\end{algorithmic}
\end{algorithm}

\newpage

\subsection*{Maximizing power for the weighted MWU statistic}

The power expression for the weighted MWU statistic is derived below. Equation (1) follows because the vector most correlated with $\delta$ is $\delta$ itself.

\begin{equation}
\begin{split}
a=\underset{a}{\operatorname{argmax}}\left(P_{H_1}\left(T>Z_{1-\alpha}\right)\right)
\\=\underset{a}{\operatorname{argmax}}\left(\Phi\left(-z_{1-\alpha}+\frac{a^T \delta}{a^T a}\right)\right)
\\=\operatorname{argmax}\left(\frac{a^T \delta}{a^T a}\right)\\ \stackrel{1}{=} \delta
\end{split}
\label{equasion:WMWU}
\end{equation}

\subsection*{Illustration of federated table construction}

The example in table~\ref{tab:algorithm_table} illustrates the algorithm with $K=10$. In the first table we see current bins with limits $0<2<4<6<8$ and columns $fx,fy$ from binning the data at the new center with those limits.

In the top panel ~\ref{tab:algorithm_table} the new center has at least 20 observations
between 4 and 6 in both groups and can potentially be divided into two new bins with privacy preserving counts that are closer to
10. In this
case we apply algorithm~\ref{alg:BinningAlgorithm} to the observations from the new center and succeed in creating new bins as
illustrated in the middle table of ~\ref{tab:algorithm_table}. The new bin limit is found from the new center's data, in exactly the same way as the bin limits were found for the first center. Splitting $(4,6]$ into two bins forces us to
reallocate the previous frequencies, 10 and 14. We do so using the relative frequencies of the new center.
The relative frequencies are $\frac{10}{20}=\frac{1}{2}$ for each bin so we allot $10\cdot\frac{1}{2}=5$ to both
new bins in group $x$ and $14\cdot\frac{1}{2}=7$ to both bins in group $y$.

\begin{table}[H]
    \centering
    
% \begin{tabular}{lrrrr}
% \toprule
% {} & \multicolumn{2}{l}{center1} & \multicolumn{2}{l}{center2} \\
% {} &             $fx$ & $fy$ &    $fx$ &  $fy$ \\
% \midrule
% (1.5,2.5] &       10 &    10 &       8 &    10 \\
% (2.5,3.5] &       15 &    20 &       20 &    20 \\
% \bottomrule
% \end{tabular}

\begin{tabular}{lrrrr}
\toprule
{} & \multicolumn{2}{l}{center1} & \multicolumn{2}{l}{center2} \\
{} &             $fx$ & $fy$ &    $fx$ &  $fy$ \\
\midrule
(0,2] & 12 & 10 & 3 & 4 \\
(2,4] & 10 & 10 & 7 & 10 \\
(4,6] & 10 & 14 & 20 & 20 \\
(6,8] & 13 & 13 & 10 & 10 \\
\bottomrule
\end{tabular}

.

\begin{tabular}{lrrrr}
\toprule
{} & \multicolumn{2}{l}{center1} & \multicolumn{2}{l}{center2} \\
{} &             $fx$ & $fy$ &    $fx$ &  $fy$ \\
\midrule
(0,2] & 12 & 10 & 3 & 4 \\
(2,4] & 10 & 10 & 7 & 10 \\
(4,5] & 5 & 7 & 10 & 10 \\
(5,6] & 5 & 7 & 10 & 10 \\
(6,8] & 13 & 13 & 10 & 10 \\
\bottomrule
\end{tabular}

.

\begin{tabular}{lrrrr}
\toprule
{} & \multicolumn{2}{l}{center1} & \multicolumn{2}{l}{center2} \\
{} &             $fx$ & $fy$ &    $fx$ &  $fy$ \\
\midrule
(0,2] & 12 & 10 & 5.46 & 7 \\
(2,4] & 10 & 10 & 4.54 & 7 \\
(4,5] & 5 & 7 & 10 & 10 \\
(5,6] & 5 & 7 & 10 & 10 \\
(6,8] & 13 & 13 & 10 & 10 \\
\bottomrule
\end{tabular}
% \begin{tabular}{lrrrr}
% \toprule
% {} & \multicolumn{2}{r}{center_1} & \multicolumn{2}{r}{center_2} \\
% group & fx & fy & fx & fy \\
% bin &  &  &  &  \\
% (1,3] & 10 & 10 & 7 & 10 \\
% (2,4] & 10 & 14 & 20 & 20 \\
% (4,6] & 13 & 13 & 10 & 10 \\
% \end{tabular}
    \caption{ Illustration of the algorithm for joining a new center}
    \label{tab:algorithm_table}
\end{table}

After creating new bins wherever possible, we iterate again and fix
bins containing non-private frequencies that are less than $K$ and have non-zero counts. This is
illustrated in table~\ref{tab:algorithm_table}. The middle table, obtained after splitting the bin  $(4,6]$, contains non-private counts from the new center: 3,7 in $(0,2],(2,4]$ from $x$ and 4 in $(0,2]$ from $y$.
This is solved in the bottom table of ~\ref{tab:algorithm_table} by
summing the frequencies in both cells and then using the relative frequencies from the current table to reallocate them. The resulting additions from the new center to the two bins are
$(3+7)\cdot\frac{12}{10+12}=5.46$ for $(0,2]$ and $(3+7)\cdot\frac{10}{10+12}=4.54$  for  $(2,4]$ in $x$ and in the same manner
7,7 for $(0,2],(2,4]$ in $y$. Note that the ``count'' added to an existing cell will not necessarily be an integer.

\section*{Funding}

This research has received funding from the European Union’s Horizon 2020 Framework Programme for Research and Innovation under the Specific Grant Agreement No. 785907 (Human Brain Project SGA2) and the Specific Grant Agreement No. 945539 (Human Brain Project SGA3).

\section*{Data Availability Statement}

The code for generating the simulation results shown in this paper can be found at https://github.com/oribech/federated-summary-table

\section*{Supplementary Material}

\subsection*{Hypothesis Testing Results}

The following table compares the combined data Mann-Whitney test and several federated alternatives. The comparison is via the 25'th and 50'th quantiles of their p-value distributions for settings where the null hypothesis of identical distributions is false. Therefore, lower p-values indicate more powerful tests.

\begin{table}[H]
    \scalebox{0.8}{
        
\begin{tabular}{llrrrr}
\toprule
   & Quantile & \multicolumn{2}{l}{0.5} & \multicolumn{2}{l}{0.25} \\
   &  Params ($\delta,\sigma_\alpha,\sigma_\beta$) & (0.063, 0.2, 0.0) & (0.063, 0.1, 0.1) & (0.063, 0.2, 0.0) & (0.063, 0.1, 0.1) \\
ncenters & Method &                   &                   &                   &                   \\
\midrule
3  & Combined  &            0.0474 &            0.0464 &            0.0101 &            0.0018 \\
   & Federated  &            0.0480 &            0.0473 &            0.0101 &            0.0018 \\
   & Fisher  &            0.0544 &            0.0161 &            0.0119 &            0.0005 \\
   & Sum  &            0.0528 &            0.0562 &            0.0111 &            0.0016 \\
   & Weighted  &            0.0463 &            0.0463 &            0.0096 &            0.0017 \\
5  & Combined  &            0.0482 &            0.0535 &            0.0106 &            0.0025 \\
   & Federated  &            0.0501 &            0.0543 &            0.0103 &            0.0026 \\
   & Fisher  &            0.0604 &            0.0191 &            0.0132 &            0.0008 \\
   & Sum  &            0.0542 &            0.0595 &            0.0119 &            0.0028 \\
   & Weighted  &            0.0452 &            0.0522 &            0.0095 &            0.0024 \\
10 & Combined  &            0.0504 &            0.0472 &            0.0101 &            0.0046 \\
   & Federated  &            0.0511 &            0.0484 &            0.0105 &            0.0048 \\
   & Fisher  &            0.0635 &            0.0227 &            0.0153 &            0.0018 \\
   & Sum  &            0.0587 &            0.0621 &            0.0122 &            0.0057 \\
   & Weighted  &            0.0480 &            0.0471 &            0.0091 &            0.0044 \\
\bottomrule
\end{tabular}

    }
    \caption{Comparison of algorithms via quantiles of their p-value distributions.}
    \label{tab:pvalues_variablility}

\end{table}

\subsection*{Quantile Estimation Results}

The tables below relate to the problem of federated estimation of quantiles with data from a Gamma distribution (shape parameter $r=4$ or $r=10$). The quantiles considered are $Q_{0.02}$, $Q_{0.25}$, $Q_{0.5}$, $Q_{0.75}$ and $Q_{0.98}$. All results relate to standardized errors  $\left( \hat{Q}_p - Q_p \right)/\sqrt{r}$. The division by $\sqrt{r}$ adjusts for the standard deviation of the underlying Gamma distribution.

Table ~\ref{tab:bias} presents the estimated bias $b$ of the various quantile estimators across the different
simulation settings.
The bias is estimated by the average of the simulation errors $\hat{b} =\frac{1}{N}\sum_{i=1}^N \left( \hat{Q}_{ip} - Q_p \right)/\sqrt{r}$.

Table~\ref{tab:sd} shows the estimated standard deviation of the estimator (after adjusting by $\sqrt{r}$), which is the sample standard deviation of the standardized errors from the simulation results.

Table~\ref{tab:err} gives the estimated mean squared error (MSE) for each estimator and each quantile, after adusting by $\sqrt{r}$. These are the square of the estimated bias plus the square of the estimated standard deviation.
Table~\ref{tab:bias_sd_ratio} presents the ratio of the bias squared to
the variance. The ratio is useful in determining whether the MSE
is dominated primarily by high variance or high bias
For most settings and values the ratio is less than 0.1, indicating that the major problem is variance.
The most notable exception is the YJ table method for estimating $Q_{0.02}$ with $r=4$, the only cases where the ratio exceeds 1. The YJ data estimator has ratios around 0.3 for many of the quantiles (except the median) when $r=4$. Again with $r=4$, both YJ estimators have ratios around 0.3 for estimating $Q_{0.98}$.
The interpolation/extrapolation estimator has high variance, and as a result, high ratios for
$Q_{0.98}$ when there are 10 centers.

\begin{table}[h]
    \scalebox{1}{
        \begin{tabular}{lllrrrrr}
\toprule
 &  &  & \multicolumn{5}{l}{Bias} \\
 &  & percentile & 0.02 & 0.25 & 0.50 & 0.75 & 0.98 \\
shape & number of centers & Method &  &  &  &  &  \\
\midrule
4 & 3 & Quantile loss & 0.001 & -0.000 & -0.001 & -0.002 & -0.002 \\
 &  & YJ data & -0.007 & 0.011 & -0.005 & -0.022 & 0.053 \\
 &  & YJ table & -0.022 & 0.006 & -0.005 & -0.016 & 0.064 \\
 & 5 & Quantile loss & 0.001 & -0.000 & 0.000 & 0.001 & 0.002 \\
 &  & YJ data & -0.008 & 0.012 & -0.004 & -0.020 & 0.061 \\
 &  & YJ table & -0.029 & 0.004 & -0.005 & -0.014 & 0.064 \\
 & 10 & Quantile loss & 0.001 & -0.000 & 0.001 & 0.002 & 0.003 \\
 &  & YJ data & -0.008 & 0.012 & -0.004 & -0.020 & 0.061 \\
 &  & YJ table & -0.036 & -0.002 & -0.008 & -0.013 & 0.088 \\
10 & 3 & Quantile loss & -0.000 & -0.000 & 0.000 & -0.000 & -0.001 \\
 &  & YJ data & -0.002 & 0.003 & -0.001 & -0.005 & 0.009 \\
 &  & YJ table & -0.013 & -0.000 & -0.002 & -0.005 & 0.004 \\
 & 5 & Quantile loss & 0.001 & -0.001 & -0.000 & -0.000 & 0.001 \\
 &  & YJ data & -0.002 & 0.003 & -0.002 & -0.005 & 0.011 \\
 &  & YJ table & -0.017 & -0.004 & -0.004 & -0.006 & 0.008 \\
 & 10 & Quantile loss & 0.002 & 0.001 & 0.000 & -0.001 & 0.001 \\
 &  & YJ data & -0.000 & 0.004 & -0.001 & -0.005 & 0.009 \\
 &  & YJ table & -0.024 & -0.007 & -0.007 & -0.007 & 0.010 \\
\bottomrule
\end{tabular}

    }
    \caption{
        Bias: the estimated bias of the quantile estimators across the different
        simulation settings.
    }
    \label{tab:bias}

\end{table}

\begin{table}[h]

    \scalebox{1}{
        
\resizebox{15cm}{!}{
\begin{tabular}{lllrrrrr}
\toprule
 &  &  & \multicolumn{5}{l}{SD} \\
 &  & percentile & 0.02 & 0.25 & 0.5 & 0.75 & 0.98 \\
shape & number of centers & Method &  &  &  &  &  \\
\midrule
%4 & 3 & Combined & 0.028 & 0.026 & 0.031 & 0.042 & 0.130 \\
% &  & Int/Ext & 0.030 & 0.026 & 0.031 & 0.042 & 0.168 \\
4 & 3 & Quantile loss & 0.028 & 0.026 & 0.031 & 0.042 & 0.130 \\
 &  & YJ data & 0.022 & 0.021 & 0.026 & 0.034 & 0.099 \\
 &  & YJ table & 0.025 & 0.021 & 0.027 & 0.035 & 0.113 \\
% & 5 & Combined & 0.029 & 0.026 & 0.031 & 0.042 & 0.129 \\
% &  & Int/Ext & 0.039 & 0.026 & 0.032 & 0.042 & 0.195 \\
 & 5 & Quantile loss & 0.029 & 0.026 & 0.031 & 0.042 & 0.129 \\
 &  & YJ data & 0.022 & 0.022 & 0.026 & 0.034 & 0.102 \\
 &  & YJ table & 0.028 & 0.022 & 0.027 & 0.035 & 0.126 \\
% & 10 & Combined & 0.028 & 0.026 & 0.031 & 0.042 & 0.126 \\
% &  & Int/Ext & 0.052 & 0.027 & 0.032 & 0.043 & 0.246 \\
 &  10& Quantile loss & 0.028 & 0.026 & 0.031 & 0.042 & 0.126 \\
 &  & YJ data & 0.022 & 0.021 & 0.027 & 0.035 & 0.100 \\
 &  & YJ table & 0.031 & 0.022 & 0.028 & 0.036 & 0.145 \\
%10 & 3 & Combined & 0.043 & 0.029 & 0.032 & 0.040 & 0.108 \\
% &  & Int/Ext & 0.047 & 0.029 & 0.032 & 0.041 & 0.140 \\
 10 & 3 & Quantile loss & 0.043 & 0.029 & 0.032 & 0.041 & 0.109 \\
 &  & YJ data & 0.033 & 0.024 & 0.027 & 0.033 & 0.082 \\
 &  & YJ table & 0.038 & 0.025 & 0.028 & 0.034 & 0.094 \\
% & 5 & Combined & 0.043 & 0.030 & 0.032 & 0.040 & 0.106 \\
% &  & Int/Ext & 0.059 & 0.030 & 0.032 & 0.040 & 0.160 \\
 &  5& Quantile loss & 0.043 & 0.030 & 0.032 & 0.040 & 0.107 \\
 &  & YJ data & 0.033 & 0.025 & 0.028 & 0.033 & 0.082 \\
 &  & YJ table & 0.041 & 0.026 & 0.029 & 0.034 & 0.101 \\
% & 10 & Combined & 0.042 & 0.030 & 0.032 & 0.039 & 0.108 \\
% &  & Int/Ext & 0.075 & 0.031 & 0.032 & 0.040 & 0.201 \\
 & 20 & Quantile loss & 0.042 & 0.030 & 0.032 & 0.039 & 0.108 \\
 &  & YJ data & 0.033 & 0.025 & 0.027 & 0.033 & 0.083 \\
 &  & YJ table & 0.048 & 0.026 & 0.028 & 0.034 & 0.120 \\
\bottomrule
\end{tabular}

}
    }
    \caption{
        SD: the estimated standard deviation of the various quantile estimators across the different
        simulation settings.
    }
    \label{tab:sd}

\end{table}

\begin{table}[h]

    \scalebox{1}{
        \resizebox{15cm}{!}{
\begin{tabular}{lllrrrrr}
\toprule
 &  &  & \multicolumn{5}{l}{MSE} \\
 &  & percentile & 0.02 & 0.25 & 0.5 & 0.75 & 0.98 \\
shape & number of centers & Method &  &  &  &  &  \\
\midrule
%4 & 3 & Combined & 0.0008 & 0.0007 & 0.0009 & 0.0018 & 0.0172 \\
% &  & Int/Ext & 0.0010 & 0.0007 & 0.0009 & 0.0017 & 0.0294 \\
 4&  3& Quantile loss & 0.0008 & 0.0007 & 0.0009 & 0.0018 & 0.0170 \\
 &  & YJ data & 0.0005 & 0.0006 & 0.0007 & 0.0017 & 0.0126 \\
 &  & YJ table & 0.0011 & 0.0005 & 0.0007 & 0.0015 & 0.0168 \\
% & 5 & Combined & 0.0008 & 0.0007 & 0.0010 & 0.0017 & 0.0166 \\
% &  & Int/Ext & 0.0016 & 0.0007 & 0.0010 & 0.0018 & 0.0387 \\
 &  5& Quantile loss & 0.0008 & 0.0007 & 0.0010 & 0.0017 & 0.0167 \\
 &  & YJ data & 0.0006 & 0.0006 & 0.0007 & 0.0016 & 0.0142 \\
 &  & YJ table & 0.0016 & 0.0005 & 0.0007 & 0.0014 & 0.0200 \\
% & 10 & Combined & 0.0008 & 0.0007 & 0.0010 & 0.0018 & 0.0159 \\
% &  & Int/Ext & 0.0029 & 0.0008 & 0.0010 & 0.0019 & 0.1291 \\
 & 10 & Quantile loss & 0.0008 & 0.0007 & 0.0010 & 0.0018 & 0.0160 \\
 &  & YJ data & 0.0005 & 0.0006 & 0.0007 & 0.0016 & 0.0138 \\
 &  & YJ table & 0.0023 & 0.0005 & 0.0008 & 0.0015 & 0.0287 \\
%10 & 3 & Combined & 0.0018 & 0.0008 & 0.0010 & 0.0016 & 0.0119 \\
% &  & Int/Ext & 0.0025 & 0.0009 & 0.0010 & 0.0017 & 0.0205 \\
 10& 3 & Quantile loss & 0.0018 & 0.0008 & 0.0010 & 0.0016 & 0.0119 \\
 &  & YJ data & 0.0011 & 0.0006 & 0.0008 & 0.0011 & 0.0067 \\
 &  & YJ table & 0.0016 & 0.0006 & 0.0008 & 0.0012 & 0.0089 \\
% & 5 & Combined & 0.0019 & 0.0009 & 0.0010 & 0.0016 & 0.0114 \\
% &  & Int/Ext & 0.0036 & 0.0010 & 0.0010 & 0.0016 & 0.0262 \\
 &  5& Quantile loss & 0.0019 & 0.0009 & 0.0010 & 0.0016 & 0.0114 \\
 &  & YJ data & 0.0011 & 0.0006 & 0.0008 & 0.0011 & 0.0068 \\
 &  & YJ table & 0.0020 & 0.0007 & 0.0008 & 0.0012 & 0.0103 \\
% & 10 & Combined & 0.0018 & 0.0009 & 0.0010 & 0.0016 & 0.0117 \\
% &  & Int/Ext & 0.0065 & 0.0010 & 0.0011 & 0.0017 & 0.0926 \\
 &  10 & Quantile loss & 0.0018 & 0.0009 & 0.0010 & 0.0016 & 0.0116 \\
 &  & YJ data & 0.0011 & 0.0006 & 0.0008 & 0.0011 & 0.0070 \\
 &  & YJ table & 0.0028 & 0.0007 & 0.0009 & 0.0012 & 0.0146 \\
\bottomrule
\end{tabular}

}
    }
    \caption{
        Error: the estimated MSE of the various quantile estimators across the different
        simulation settings.
    }
    \label{tab:err}

\end{table}

\begin{table}[h]

    \scalebox{1}{
        \resizebox{15cm}{!}{
\begin{tabular}{lllrrrrr}
\toprule
 &  &  & \multicolumn{5}{l}{Bias2/Var} \\
 &  & percentile & 0.02 & 0.25 & 0.5 & 0.75 & 0.98 \\
shape & number of centers & Method &  &  &  &  &  \\
\midrule
4 & 3 & Quantile loss & 0.0013 & 0.0000 & 0.0011 & 0.0023 & 0.0002 \\
 &  & YJ data & 0.1025 & 0.2676 & 0.0360 & 0.4117 & 0.2877 \\
 &  & YJ table & 0.7513 & 0.0779 & 0.0351 & 0.2124 & 0.3209 \\
 & 5 & Quantile loss & 0.0012 & 0.0000 & 0.0000 & 0.0006 & 0.0002 \\
 &  & YJ data & 0.1269 & 0.3114 & 0.0229 & 0.3388 & 0.3550 \\
 &  & YJ table & 1.0782 & 0.0334 & 0.0349 & 0.1587 & 0.2577 \\
 & 10 & Quantile loss & 0.0013 & 0.0000 & 0.0010 & 0.0023 & 0.0006 \\
 &  & YJ data & 0.1383 & 0.3145 & 0.0223 & 0.3306 & 0.3708 \\
 &  & YJ table & 1.3263 & 0.0080 & 0.0841 & 0.1272 & 0.3678 \\
10 & 3 & Quantile loss & 0.0000 & 0.0000 & 0.0000 & 0.0000 & 0.0001 \\
 &  & YJ data & 0.0037 & 0.0153 & 0.0013 & 0.0226 & 0.0122 \\
 &  & YJ table & 0.1168 & 0.0000 & 0.0051 & 0.0221 & 0.0018 \\
 & 5 & Quantile loss & 0.0005 & 0.0011 & 0.0000 & 0.0000 & 0.0001 \\
 &  & YJ data & 0.0036 & 0.0146 & 0.0052 & 0.0225 & 0.0181 \\
 &  & YJ table & 0.1687 & 0.0246 & 0.0196 & 0.0316 & 0.0063 \\
 & 10 & Quantile loss & 0.0023 & 0.0011 & 0.0000 & 0.0006 & 0.0001 \\
 &  & YJ data & 0.0000 & 0.0257 & 0.0013 & 0.0235 & 0.0117 \\
 &  & YJ table & 0.2552 & 0.0741 & 0.0605 & 0.0423 & 0.0069 \\
\bottomrule
\end{tabular}
}

    }
    \caption{
        Ratio: the estimated ratio between the squared bias and the variance
        for the quantile estimators across the different simulation settings.
    }
    \label{tab:bias_sd_ratio}

\end{table}

\end{document}